\documentclass[aip, twocolumn,nofootinbib, floatfix, superscriptaddress, reprint, 10pt]{revtex4-1}
\newcommand{\bA}{\mathbf{A}}
\newcommand{\bAhat}{\hat{\mathbf{A}}}

\newcommand{\bcol}[2]{{#1}_{#2}}
\newcommand{\brow}[2]{{#1}_{#2}}

\newcommand{\opd}{\operatorname{d}}
\newcommand{\dt}{\tilde{\opd}}

\newcommand{\nf}{n_{\text{features}}}
\newcommand{\ns}{n_{\text{samples}}}
\newcommand{\np}{n_{\text{properties}}}
\newcommand{\npca}{n_{\text{latent}}}

\newcommand{\bP}[2]{\mathbf{P}_{{#1}{#2}}}

\newcommand{\bPtx}{\bP{T}{X}}
\newcommand{\bPty}{\bP{T}{Y}}

\newcommand{\bPxt}{\bP{X}{T}}

\newcommand{\bX}{\mathbf{X}}
\newcommand{\bXcol}[1]{\bcol{\bX}{#1}}
\newcommand{\bXrow}[1]{\brow{\bx}{#1}}
\newcommand{\lvague}{*}
\newcommand{\lc}{c}
\newcommand{\lr}{r}
\newcommand{\bvague}{\scalebox{1.25}{$\boldsymbol{\lvague}$}}
\newcommand{\bc}{\mathbf{\lc}}
\newcommand{\br}{\mathbf{\lr}}

\newcommand{\bXf}{\bXcol{\bc}}
\newcommand{\bXft}{\bcol{\tilde{\bX}}{\bc}}
\newcommand{\bXs}{\bXcol{\br}}
\newcommand{\bXst}{\bcol{\tilde{\bX}}{\br}}

\newcommand{\bYs}{\brow{\bY}{\br}}

\newcommand{\bXt}{{\tilde{\bX}}}
\newcommand{\bXhat}{{\hat{\bX}}}

\newcommand{\bx}{{\mathbf{x}}}

\newcommand{\bY}{\mathbf{Y}}

\newcommand{\by}{\mathbf{y}}
\newcommand{\bYhat}{\hat{\bY}}
\newcommand{\byhat}{\hat{\by}}

\newcommand{\bKNN}{\bG}
\newcommand{\bKNM}{\bG_{\br}}
\newcommand{\bKMM}{\bG_{\br\br}}

\newcommand{\bI}{\mathbf{I}}

\newcommand{\bS}{\left(\bXf^- \bX \bXs^-\right)}
\newcommand{\bC}{\mathbf{C}}
\newcommand{\bCt}{\tilde{\mathbf{C}}}

\newcommand{\bT}{\mathbf{T}}

\newcommand{\bG}{\mathbf{K}}
\newcommand{\bGt}{\tilde{\bG}}

\newcommand{\bU}{\mathbf{U}}
\newcommand{\bUhat}{\hat{\bU}}

\newcommand{\bLAM}{\boldsymbol{\Lambda}}
\newcommand{\bLAMhat}{\hat{\bLAM}}

\newcommand{\bSIG}{\bLAM^{1/2}}

\newcommand{\loss}[2]{\frac{\lVert {#1} - {#2} \rVert^2}{\lVert{#1}\rVert^2}}
\newcommand{\lproj}{\loss{\bX}{\bT\bPtx}}

\newcommand{\lregr}{\loss{\bY}{\bT\bPty}}

\newcommand{\eg}{\textit{e.g.}~}
\newcommand{\legendexp}[1]{Metrics shown are PCov-CUR (red), PCov-FPS with random {#1}$_0$ (blue), and random feature selection (black, solid). Shaded regions bound the minimum and maximum error across all $\alpha$ for each PCovR-inspired method. }
\newcommand{\gd}[1]{the grey dashed line denotes the error using the full {#1}.}

\usepackage{bm}
\usepackage{graphicx}
\usepackage[usenames]{color}
\usepackage{microtype}
\usepackage{amsmath}
\usepackage{amssymb}
\usepackage{xspace}
\usepackage{comment}
\usepackage{footnote}
\usepackage[version=4]{mhchem}
\usepackage{subcaption}
\usepackage{relsize}
\usepackage{svg}
\usepackage{hhline}
\usepackage{verbatim}
\usepackage[normalem]{ulem}
\usepackage[inline]{enumitem}
\usepackage{lipsum}
\usepackage{xr}

\makeatletter
\newcommand*{\addFileDependency}[1]{%
  \typeout{(#1)}
  \@addtofilelist{#1}
  \IfFileExists{#1}{}{\typeout{No file #1.}}
}
\makeatother

\newcommand*{\myexternaldocument}[1]{%
    \externaldocument{#1}%
    \addFileDependency{#1.tex}%
    \addFileDependency{#1.aux}%
}
\myexternaldocument{SI/SI}

\definecolor{mygray}{gray}{0.5}

\makeatletter

\renewcommand*{\p@subsection}{}

\renewcommand*{\p@subsubsection}{}

\renewcommand*{\p@paragraph}{}
\renewcommand*{\p@figure}{Fig.~}
\makeatother

\begin{document}

\title{Improving Sample and Feature Selection with Principal Covariates Regression}

\author{Rose K. Cersonsky}
\affiliation{Laboratory of Computational Science and Modeling, IMX, \'Ecole Polytechnique F\'ed\'erale de Lausanne, 1015 Lausanne, Switzerland}

\author{Benjamin A. Helfrecht}
\affiliation{Laboratory of Computational Science and Modeling, IMX, \'Ecole Polytechnique F\'ed\'erale de Lausanne, 1015 Lausanne, Switzerland}

\author{Edgar A. Engel}
\affiliation{TCM Group, Cavendish Laboratory, University of Cambridge, J.J. Thomson Avenue, Cambridge CB3 0HE, United Kingdom}

\author{Michele Ceriotti}
\email{michele.ceriotti@epfl.ch}
\affiliation{Laboratory of Computational Science and Modeling, IMX, \'Ecole Polytechnique F\'ed\'erale de Lausanne, 1015 Lausanne, Switzerland}
\begin{abstract}
Selecting the most relevant features and samples out of a large set of candidates is a task that occurs very often in the context of automated data analysis, where it can be used to improve the computational performance, and also often the transferability, of a model. 
Here we focus on two popular sub-selection schemes which have been applied to this end: CUR decomposition, that is based on a low-rank approximation of the feature matrix and Farthest Point Sampling, that relies on the iterative identification of the most diverse samples and discriminating features. 
We modify these unsupervised approaches, incorporating a supervised component following the same spirit as the Principal Covariates Regression (PCovR) method.  We show that incorporating target information provides selections that perform better in supervised tasks, which we demonstrate with ridge regression, kernel ridge regression, and sparse kernel regression. We also show that incorporating aspects of simple supervised learning models can improve the accuracy of more complex models, such as feed-forward neural networks. 
We present adjustments to minimize the impact that any subselection may incur when performing unsupervised tasks. 
We demonstrate the significant improvements associated with the use of PCov-CUR and PCov-FPS selections for applications to chemistry and materials science, typically reducing by a factor of two the number of features and samples which are required to achieve a given level of regression accuracy.
\end{abstract}

\maketitle

\section{Introduction}

In recent years, machine learning (ML) models have found application across a vast breadth of scientific fields, from economics \cite{intro-econ-Bolton2002, intro-econ-Fischer2018, intro-econ-Huang2004, intro-econ-Tsai2008} to medical diagnostics \cite{intro-diag-Guyon2002, intro-diag-Peng2019, intro-diag-Rajkomar2018, intro-diag-Wolf2018} to sensing \cite{intro-sens-Belgiu2016, intro-sens-Gramfort2014, intro-sens-Mountrakis2011} to computational chemistry\cite{intro-cc-Berrueta2007, intro-cc-Daina2017, intro-cc-McGibbon2015}. 
Even though data-driven modelling is often discussed in a ``big data'' context, where access to data is inexpensive and the computational cost of a machine learning model is of secondary importance to regression accuracy, many applications benefit greatly by a reduction of their data requirements, and/or the acceleration of training and prediction. 
The search for a balance between the complexity of the model, the amount of training data and the accuracy of predictions has given rise to a subclass of ML schemes focused on data and feature sub-selection, wherein a subset of samples or descriptors is identified that minimises the corresponding degradation of accuracy relative to the full model\cite{Blum1997, Li2018}.

The objective of sample selection is to condense the sample space to only those with statistical significance, effectively pruning the redundant samples and identifying ideal candidates for costlier reference calculations or analysis steps\cite{Xu2019}.
Methods may seek to find a \textit{core-set} that is representative of the entire sample space, \eg through voronoi tessellations\cite{Du1999}, committee models\cite{Garcia-Osorio2010, akdemir2015} or random forests\cite{Wang2012} or alternatively to temper the error in representing outlier or border samples, such as  with sensitivity heuristics\cite{Widrow1960, Zeng2001, Ng2003} or nearest neighbour analysis \cite{Hart1968}.
Conversely, in feature selection one determines an information-rich subset of large and/or sparse features. While this motivation is akin to traditional dimensionality reduction techniques like Principal Components Analysis (PCA), which construct \textit{new} features as combinations of the input features, feature selection differs in its preservation of the original feature space. This is of particular importance where features hold conceptual value, \eg sensors for autonomous robots\cite{Balakrishnan1996}, medical markers for diagnostic classification \cite{intro-diag-Guyon2002, Ding2005, Fan2010, Chuang2008}, or words for predictive text analysis\cite{Kuhn2019, Uguz2011, Nicholls2010, Lewis1992}.

Most sub-selection methods are  \textit{unsupervised}, seeking to exploit the diversity of the selections in order to maximise the variance in the samples or features.
For example, farthest point sampling (FPS) maximises diversity of the selected vectors as measured by the mutual Euclidean distance, and selection methods based on the CUR decomposition choose the columns and/or rows of the feature matrix that allow one to construct the best low-rank approximation of the full matrix. 
However, in unsupervised selection models, preservation of pertinent information for supervised tasks is not guaranteed, particularly in the case of poor representations or non-linear relationships between features and targets.
Thus, for supervised tasks, it may be attractive to use supervised or semi-supervised selection that use knowledge of the regression target space to influence the choice of the most important samples or features.

\begin{figure*}[ht]
\centering
\includegraphics[width=\linewidth]{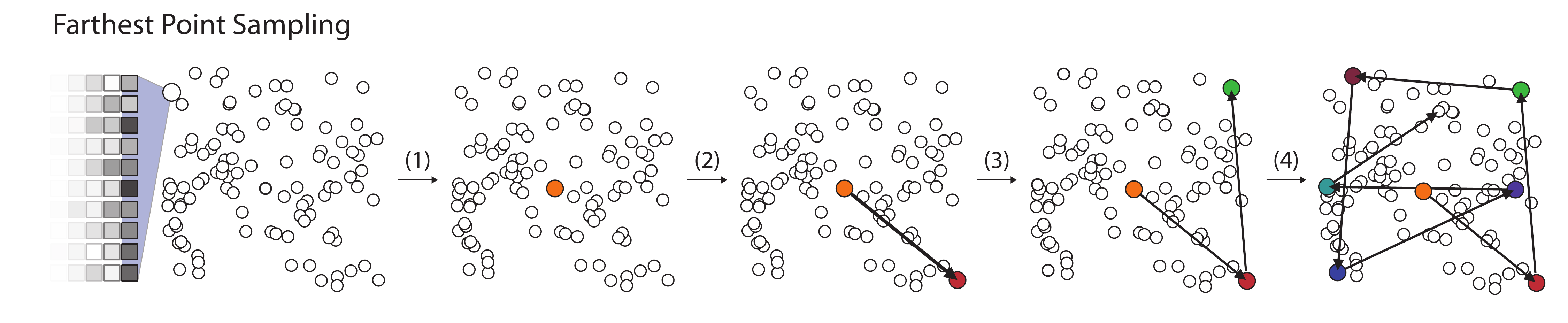}
\caption{\textbf{FPS selection.} 
Each point in the plot corresponds to a row (for sample selection) or column (for feature selection) of the matrix. \textbf{(1)} Select an initial item. \textbf{(2)} Select the item that is farthest from (1). \textbf{(3)} Select the item which is farthest from those already selected. \textbf{(4)} Repeat until the target number of items is selected.}
    \label{fig:fps}
\end{figure*}

Inspired by the Principal Covariates Regression (PCovR) method\cite{de_jong_principal_1992}, we propose a modification to the FPS and CUR approaches that combines the unsupervised component with an explicit assessment of the performance of the sub-selection in the context of property regression tasks. 
We demonstrate the superior performance of these algorithms to select features or samples for supervised learning and neural networks, and discuss small modifications that improve the performance of these subsets for unsupervised tasks. 
As a representative application of these methods we discuss the atomic-scale study of molecules and materials, wherein the features encode structural and compositional information, that is used to predict properties such as the magnetic chemical shieldings of nuclei\cite{cuny+16jctc,paru+18ncomm}, the energy associated with an atomistic configuration or the forces acting on its atoms\cite{behl-parr07prl,bart+10prl,rupp+12prl}.

\section{Methods}

In discussing the feature and sample selection methods we introduce here, we assume that the reader is familiar with simple linear and kernel methods; those unacquainted with these methods or wishing further explanation may refer to \citet{helf+20mlst}, which contains a pedagogic discussion of these methods in a notation similar to that used below.

\subsection{Notation}

\paragraph {$\bX$ and $\bY$} For each system, we will assume that inputs are described by a $\ns \times \nf$ matrix $\bX$, where each row vector $\bx$ contains as its entries the features of the corresponding sample. We will also assume that the $\ns \times \np$ matrix $\bY$ consists of rows (denoted $\by$) containing the properties corresponding to the samples in $\bX$. Furthermore, we will assume that $\bX$ and $\bY$ have been centred by their column means and scaled such that $\bX$ has unit variance and each column of $\bY$ has variance equal to $(1/\np)$. The standardisation step is not essential, but ensures that features and properties are a-dimensional and have a natural variability of the order of 1. 

\paragraph {Projectors and Latent Space} We will use $\bT$ to denote a projection of data into a lower-dimensional latent space. We will also use $\bP{A}{B}$ to denote a projector from one space to another (here from the space of $\bA$ to the space of $\mathbf{B}$). For instance, in this notation a linear projection from feature to latent space reads $\bT = \bX \bPxt$.

\paragraph{Matrix slices} For a general matrix $\bA$, we will denote a general subset of the elements of $\bA$ as $\bA_{\bvague}$. The feature-selected subset of $\bA$ is given as $\bcol{\bA}{\bc}$, consisting of the $M$ features found in columns $\bc = (\lc_0, \lc_1, \lc_2, ... \lc_M)$. The sample-selected subset of $\bA$ is given as $\brow{\bA}{\br}$, consisting of the $M$ samples found in rows $\br = (\lr_0, \lr_1, \lr_2, ... \lr_M)$. We indicate the $i^{th}$ row  as $\mathbf{a}_i$ and the $j^{th}$ column as $\bA_j$, while the $j^{th}$ element  in the $i^{th}$ row is given by $A_{ij}$.

\paragraph{Accents and Operations} We will use $\bAhat$ to indicate an approximation of $\bA$ and $\tilde{\bA}$ for an augmentation (a matrix that is conceptually analogous to $\bA$, but is modified to incorporate different kinds of information).
We will also use $\bU_A$ and $\bLAM_A$ to represent the eigenvectors and eigenvalues of a matrix $\bA$ = $\bU_A \bLAM_A \bU_A^T$, where $\bU_A$ contains the eigenvectors as columns. We will denote the pseudoinverse of a non-invertible matrix $\bA^-$, which is equal to $\left(\bA^T \bA\right)^{-1}\bA^T$.

\paragraph{Loss Measures}
We will report different loss measures, where $\ell_{A}$ represents the relative loss in reconstructing $\bA$ given the corresponding method, where generally $$\ell_{A} = \loss{\bA}{\bAhat}.$$ We will often omit the denominator, given our choice of standardising feature and property matrices.
Where appropriate, we will use the root mean-squared error (RMSE) to give regression losses in concrete units. 

\subsection{Selection Methods}

Selecting samples or features amounts to picking rows and columns of $\bX$ that provide model performances comparable to that of the full feature matrix.
Of the many strategies that have been proposed to perform this selection, we will build upon two:
Farthest Point Sampling (FPS) and CUR Decomposition, whose functioning we summarise in this Section.

\begin{figure*}
    \centering
    \includegraphics[width=\linewidth]{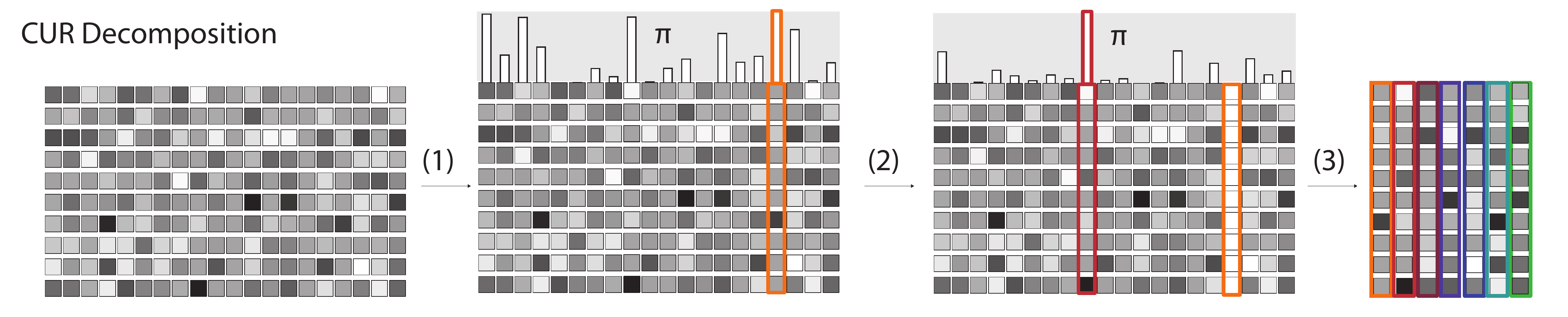}
    \caption{\textbf{CUR selection.}
    The figure demonstrates CUR \textit{feature} selection. Sample selection proceeds identically with the matrix rows used in place of the columns.
    \textbf{(1)} Compute the importance score $\pi$ for each column; select the feature that maximises $\pi$. \textbf{(2)} Orthogonalise the matrix with respect to selected feature; recompute $\pi$; select the feature which maximises $\pi$; \textbf{(3)} Repeat step (2) until the target number of items is obtained. 
    }
    \label{fig:cur}
\end{figure*}

\subsubsection{Farthest Point Sampling}

Farthest Point Sampling (FPS) relies on the definition of a metric in the input space,  and uses it to maximise the diversity of the selection\cite{FPS}. FPS is a greedy selection scheme (meaning that points are selected incrementally) and it is deterministic -- except for the choice of the first point, which is usually picked at random. Each subsequent choice is made to maximise the distance to the points that have been previously selected
\begin{equation}
    \lvague_{m+1} = \operatorname{argmax}_j \left\{ \min_{i \in \bvague_m}\left[\opd(i,j)\right]\right\}
\end{equation}
where $\bvague_m$ contains the previous selections, $\lvague_{m+1}$ is the next selected sample or feature, and $\opd(i,j)$ indicates the distance between the $i^{th}$ and $j^{th}$ column or row. A schematic of this process is depicted in ~\ref{fig:fps}. 

Even though any metric can be used to define $\opd(i,j)$, one often simply uses an Euclidean distance\cite{imba+18jcp}. 
For sample selection, traditional FPS employs a row-wise Euclidean distance
\begin{equation}
\opd_\lr(i, j) = \left\| \bXrow{i} -\bXrow{j} \right\|^2.
\label{eq:fps_sample}
\end{equation}
For feature selection, %
the corresponding column-wise Euclidean distance is
\begin{equation}
\opd_\lc(i, j) = \left\| \bXcol{i} -\bXcol{j} \right\|^2.
\label{eq:fps_feature}
\end{equation}

These two metrics can be also expressed in terms of the Gram matrix $\bG = \bX \bX^T$
\begin{equation} \label{eq:fps-K}
    \opd_\lr(i, j) = K_{ii} - 2 K_{ij} + K_{jj}
\end{equation}
and  the covariance matrix $\bC = \bX ^ T \bX$
\begin{equation}\label{eq:fps-C}
    \opd_\lc(i, j) = C_{ii} - 2 C_{ij} + C_{jj},
\end{equation}
two definitions that will be useful in extending FPS to incorporate information on the properties $\bY$. 
Eq.~\eqref{eq:fps-K} is also useful to define a metric in cases in which the problem is described through a kernel formulation rather than through an explicit set of features -- which allows injecting a degree of non-linearity in the ML model by defining a kernel function $k(\bx,\bx')$ that expresses the similarity between pairs of samples\cite{scho+98nc}.
By setting $K_{ij}=k(\bx_i,\bx_j)$, one can perform sample-space FPS  using exactly the same procedure discussed here, in a way that is consistent with the kernel-induced metric. 

\begin{figure*}
    \centering
    \includegraphics[width=\linewidth]{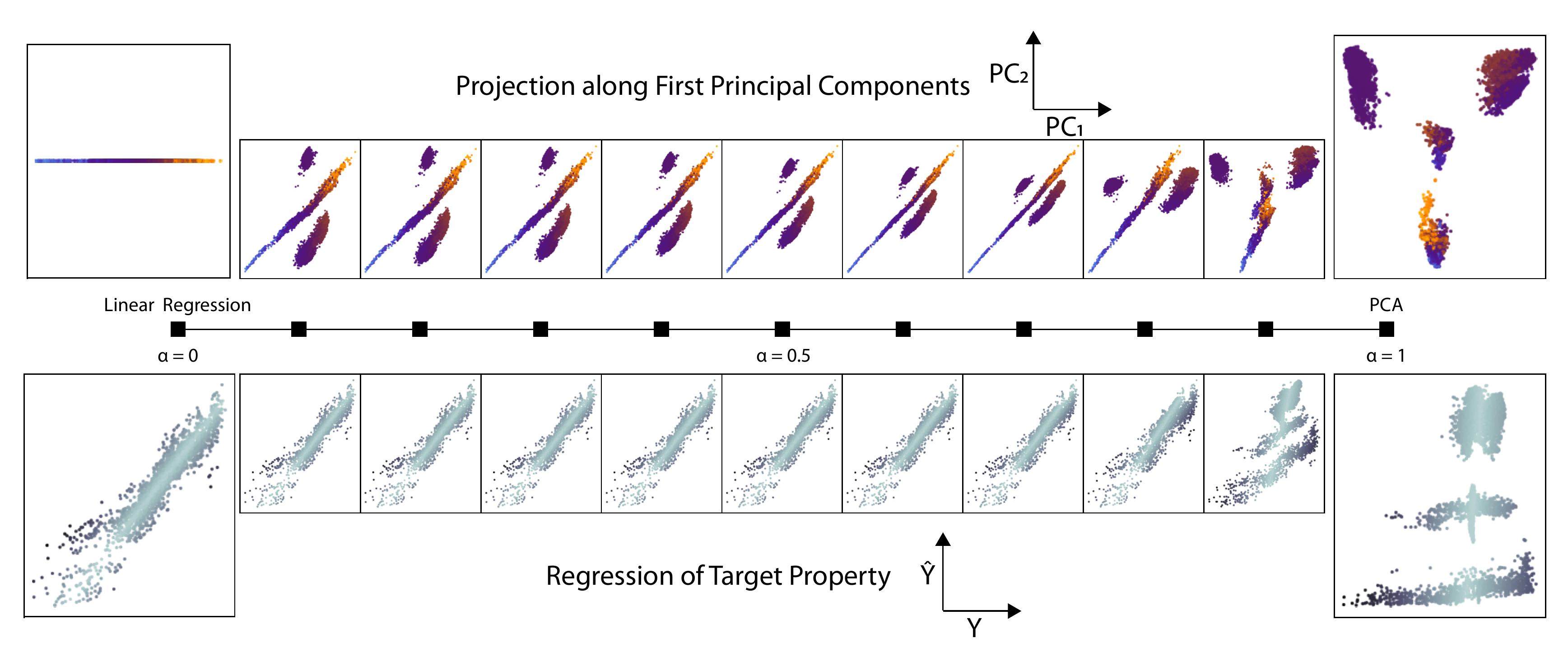}
    \caption{\textbf{Example projections and regressions using the PCovR model.} PCovR provides a projection into latent space which optimises a combined regression and projection loss, as weighted by the parameter $\alpha$, where $\alpha=0$ corresponds to pure linear regression and $\alpha=1$ corresponds to principal components analysis (PCA). Projections (top row) are given across the first two principal components and coloured by the corresponding property. $\bY$ vs $\bYhat$ regressions plots (bottom row) are coloured by regression error, with darker corresponding to higher error. As seen here, intermediate $\alpha$ provides a projection which performs better in both tasks than either linear regression or PCA.}
    \label{fig:PCovR}
\end{figure*}

\subsubsection{CUR Decomposition}
\label{sec:cur}

CUR decomposition\cite{maho-drin09pnas} aims to approximate a matrix $\bX$ using a subset of columns and rows, such that 
\begin{equation}
    \bXhat \approx \bXf \left(\bXf^- \bX \bXs^-\right) \bXs.
    \label{eq:cur}
\end{equation}
Typically in CUR, $\bXf$, $\bS$, and $\bXs$ are denoted $\bC$, $\bU$, and $\mathbf{R}$, however we have changed notation to avoid confusion with the covariance matrix $\bC$ and eigenvectors $\bU$. For a given choice of rows and columns, Eq.~\eqref{eq:cur} gives the best approximation of the original feature matrix in terms of $\bXf$ and $\bXs$. 
Various implementations of CUR only differ by the strategy chosen to select $\bc$ and $\br$.
These subsets of rows and columns are usually determined incrementally, computing at each stage a \emph{leverage score} $\pi$, representative of the relative importance of each column or row,
\begin{equation}
    *_{m+1} = \operatorname{argmax}_j\left\{\pi_j\right\}
\end{equation}
Many CUR flavors, including \citet{maho-drin09pnas}, incorporate an element of randomness in the selection -- mostly to improve performance in the limit of large data sets. In this paper, we utilise a deterministic variant, introduced in \citet{imba+18jcp}. 
In each iteration, after having selected the entry $j$ for which $\pi_j$ is highest, we orthogonalise the remaining columns or rows with respect to that which has just been selected.

In the most common form of CUR, the leverage score is computed from the singular value decomposition\cite{golub_singular_1970,klema_singular_1980} of the feature matrix, $\bX=\bU_\bG\bSIG\bU_\bC^T$. The right singular vectors of $\bX$, $\bU_\bC$, coincide with the eigenvectors of the covariance $\bC=\bX^T \bX$. The left singular vectors, $\bU_\bG$, are equal to the eigenvectors of the Gram matrix $\bG=\bX\bX^T$. $\bSIG$ is a rectangular-diagonal matrix with the singular values, which equal the square roots of the non-zero eigenvalues of either $\bC$ or $\bG$. For selecting samples, $\pi$ is the sum over the squares of the first $k$ components of the left singular vectors
\begin{equation}\label{eq:leverage-K}
    \pi_i = \sum_j^k \left(\bU_\bG\right)_{ij}^2,
\end{equation}
We orthogonalise the remaining samples after each iteration with respect to the most recently selected row $\bXrow{\lr}$
\begin{equation}
\bX \leftarrow \bX -  \bX \left(\frac{\bXrow{\lr}^T \bXrow{\lr}}{\lVert \bXrow{\lr}\rVert^2}\right).
\end{equation}
For feature selection, the score is computed with the \emph{right} singular vectors.
\begin{equation}
    \pi_i = \sum_j^k \left(\bU_\bC\right)_{ij}^2\,,
\end{equation}
and the orthogonalisation is performed relative to the most recently selected column $\bXcol{\lc}$
\begin{equation}
\bX \leftarrow \bX -  \left(\frac{\bXcol{\lc}\bXcol{\lc}^T}{\lVert\bXcol{\lc}\rVert^2}\right)\bX.
\end{equation}
The number  $k$  of singular vectors included in computing the leverage score should usually be small, and $k=1$ performs well when the deterministic procedure we discuss here is followed.

These steps are iterated until a sufficient number of features has been selected. A schematic of feature selection using CUR is given in \ref{fig:cur}.
This deterministic approach, that requires performing a new singular value decomposition at each iteration, generally outperforms the more traditional approach, wherein one selects all features in a single iteration. This is further demonstrated in Fig. S2.

\subsection{Principal Covariates Regression}
\label{sec:pcovr}

Principal covariates regression (PCovR)\cite{de_jong_principal_1992} is an algorithm used to generate a latent space projection that minimises a combined PCA and LR-like loss
\begin{equation}
    \ell = \alpha \lproj + \left(1-\alpha\right) \lregr.
\end{equation}
where $\alpha$ is a mixing parameter that determines the relative weight of the two components. Setting $\alpha=0.0$ corresponds to linear regression, and $\alpha=1.0$ corresponds to PCA.

There are two routes of obtaining this latent space projection in PCovR\cite{de_jong_principal_1992, helf+20mlst}. In \emph{sample-space PCovR},  the latent space projection is determined by the modified gram matrix of size $\ns \times \ns$:
\begin{equation}
\bGt = \alpha\bX\bX^T + (1 - \alpha)\bYhat\bYhat^T,
\label{eq:gram-pcovr}
\end{equation}
where $\bYhat$ is the result of an appropriate regression approximation of $\bY$ to avoid over-fitting. The $\ns \times \npca$ latent space projection is then given by $\bT = \bUhat_{\bGt}\bLAMhat_{\bGt}^{1/2}$,
truncated to the $\npca$ largest eigenvalues and their corresponding eigenvectors. 

However, computing the eigendecomposition $\bGt$ is intractable for a large number of samples; therefore, when $\nf < \ns$, it is advantageous to perform \emph{feature-space PCovR}, where an equivalent latent projection is determined by the eigendecomposition of a modified covariance matrix of size $\nf \times \nf$:
\begin{equation}
\bCt = \bC^{-1/2} \bX^T \bGt \bX \bC^{-1/2}
\label{eq:cov-pcovr}
\end{equation}
The $\ns \times \npca$ latent space projection is then given by $\bT = \bX {\bC}^{-1/2} \bUhat_{\bCt} {\bLAMhat_{\bCt}}^{1/2}$.

Examples of the projections and regressions obtained using PCovR, performed on the NMR Chemical Shieldings of the CSD-1000R dataset \cite{paru+18ncomm} are shown in \ref{fig:PCovR}. In the $\alpha=0.0$ case, the projection contains solely the regression weight(s), and the second principal component is zero, as this dataset has $\np=1$. %
In the $\alpha=1.0$ case, the projection distinguishes the clusters (that are associated to the chemical nature of the atoms, that can be HNCO) but fails to regress the properties. For most PCovR models, an intermediate value of $\alpha \approx 0.5$ optimises the combined loss\cite{vervloet_selection_2013, vervloet_pcovr:_2015, helf+20mlst}. 
In the many cases in which a kernel ridge regression (KRR) model outperforms plain linear regression, the performance of PCovR can be similarly improved using its kernelised counterpart, KPCovR, where $\bGt = \alpha \bG + (1-\alpha)\bYhat \bYhat^T$, with $\bG$ being the kernel matrix and $\bYhat$ the predicted properties obtained through KRR \cite{helf+20mlst}.

\subsection{Feature and Sample Selection with PCovR}

The idea of injecting information from a supervised task into an unsupervised ML scheme is particularly appealing when it comes to features and samples selection to be used for supervised tasks. 
It is relatively easy to modify FPS and CUR to incorporate a PCovR-like combination of feature and property space, by modifying the definitions of the metric and the leverage score.

\subsubsection{PCov-FPS} 

The formulation of a PCovR-inspired version of FPS for sample selection is rather straightforward, as it simply involves replacing the distance $\opd$ in Eq.~\eqref{eq:fps_sample} with an augmented distance $\dt$
\begin{equation}
\begin{split}
    \dt_{\lr}(i,j) = \alpha \lVert \bXrow{i} - \bXrow{j}\rVert^2 + (1-\alpha) \lVert \byhat_i - \byhat_{j}\rVert^2
\end{split}
\label{eq:pcov-fps-sample-dist}
\end{equation}
With this definition, the method linearly interpolates between Euclidean FPS as defined in Eq.~\ref{eq:fps_sample} at $\alpha = 1.0$ and one which maximises the diversity of $\bY$ at $\alpha=0.0$.

By writing out explicitly the distances in terms of scalar products, one can see that this definition is equivalent to Eq.~\eqref{eq:fps-K} replacing $\bG$ with the PCovR version of the Gram matrix $\bGt$
\begin{equation}\label{eq:fps-Kt}
    \dt_{\lr}(i,j) = \tilde{K}_{ii} - 2 \tilde{K}_{ij} + \tilde{K}_{jj}.
\end{equation}
The extension to KPCov-FPS is trivial, by just using the PCov extension of a kernel matrix $\bG$, as discussed in \citet{helf+20mlst}.

A feature-space version of PCov-FPS can be obtained by using a feature distance analogous to Eq.~\eqref{eq:fps-C}, computed using $\bCt$, resulting in the metric
\begin{equation}\label{eq:fps-Ct}
   \dt_{\lc}(i,j) = \tilde{C}_{ii} - 2 \tilde{C}_{i j} + \tilde{C}_{jj}.
\end{equation}

\subsubsection{PCov-CUR}

To incorporate PCovR into CUR-based selection, we propose to proceed as in Section~\ref{sec:cur}, but to compute the leverage scores using $\bU_{\bGt}$ and $\bU_{\bCt}$ in place of the left and right singular vectors. 
This is motivated by the fact that $\bCt$ and $\bGt$ share the same relationship as $\bC$ and $\bG$, and that one could define PCovR-style features $\tilde{\bX}$ whose SVD yields $\bU_{\bGt}$ and $\bU_{\bCt}$ as left and right singular vectors (this is briefly discussed in Appendix A, with the full proof in Sec. S1). 
For sample selection, we compute the eigenvectors of the  PCov-CUR gram matrix $\bGt$ (Eq.~\eqref{eq:gram-pcovr}), and for feature selection those of the PCovR covariance matrix $\bCt$ (Eq.~\eqref{eq:cov-pcovr}). We found that the best results are obtained by computing leverage scores using a number of eigenvectors $k$ that is smaller or equal than the number of properties in $\bY$, as demonstrated in Fig. S1.

The only additional change that needs to be incorporated in the procedure described in Sec.~\ref{sec:cur} involves eliminating at each step the components of the property matrix that can be described by the selected features or samples. One should perform the update
\begin{equation}
    \bY \leftarrow \bY - \bXf \left(\bXf^T \bXf\right)^{-1}\bXf^T \bY,
\end{equation}
so that the next iteration of the CUR selection identifies the features that are best suited to describe the residual error in the predicted properties.

For sample selection, the update has to be designed to remove the components of the property matrix described by the regression trained on the selected samples
\begin{equation}
    \bY \leftarrow \bY - \bX \left(\bXs^T \bXs\right)^{-1}\bXs^T \bYs.
\end{equation}

\section{Results}

CUR and FPS can be applied in the context of different ML problems. In this paper we assess the performance of their PCov-styled counterparts covering several scenarios that are common in supervised learning, although we also discuss the implications for unsupervised tasks. 
In particular, we will benchmark feature selection in the context of linear, kernel and non-linear regression problems, sample selection as a strategy to reduce the train set size, and active set selection for sparse kernel ridge regression methods. 
For every model and task we compare a fully random selection of features, PCov-CUR and PCov-FPS with different values of $\alpha$, noting that $\alpha=1$ corresponds to standard FPS and CUR.
Due to the random initialisation of (PCov-)FPS, multiple PCov-FPS selections were performed, and average errors for each $\alpha$ reported.
Unless otherwise stated, for all supervised models the hyperparameters are optimised by 2-fold cross-validation. 

\begin{figure*}[t]
    \includegraphics[width=\linewidth]{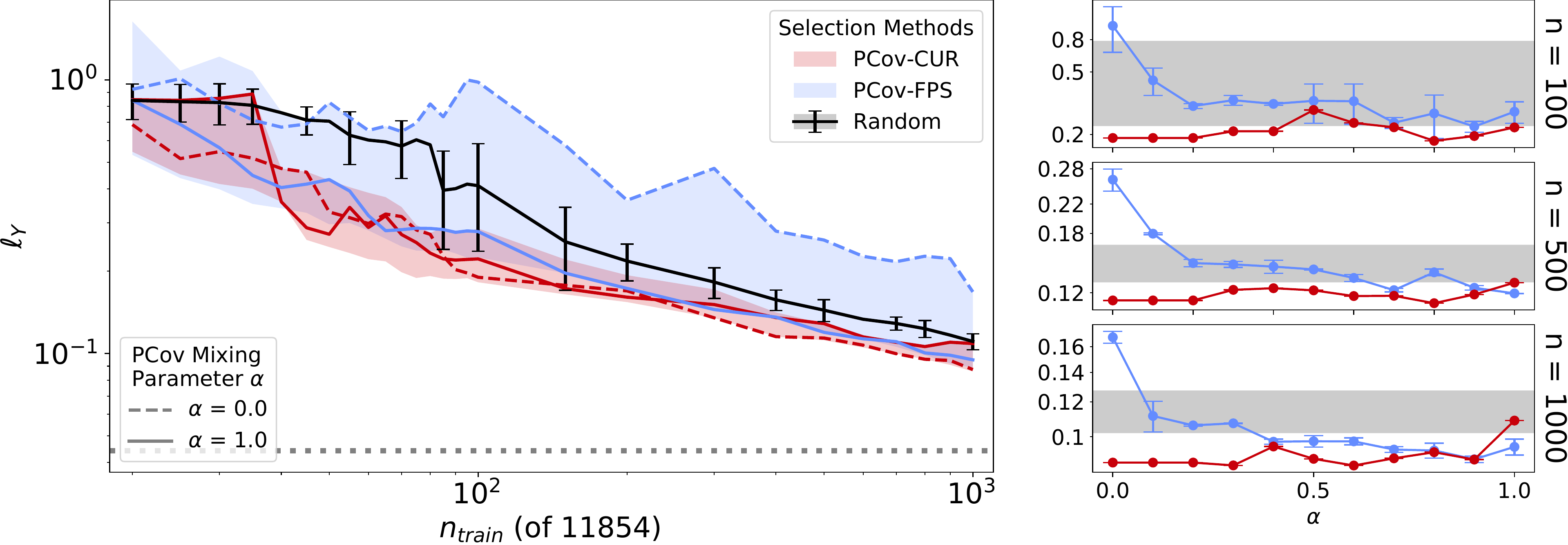}
    \caption{\textbf{Mean relative error using CSD-1000R training subsets}. \legendexp{\lr} In the left panel \gd{training set}
    The right side panels show the performance across $\alpha$ at $\ns=100, 500, 1000$  (of 11'854) samples. 
    } 
    \label{fig:ts_lr_ly}
\end{figure*}

While our approach is completely general, we focus here on a benchmark system that is relevant for applications of ML to atomistic simulations, chemistry and materials science. The CSD-1000r\cite{musi+19jctc} dataset contains \ce{C},\ce{H},\ce{N},\ce{O} atomic environments taken from 1000 crystal structures of molecular compounds, and their NMR chemical shieldings as target properties. 
We describe the atom-centred environments in terms of power spectrum SOAP vectors\cite{bart+13prb}, which have been previously employed for supervised and unsupervised tasks on similar datasets\cite{paru+18ncomm,enge+19pccp,helf+20mlst}. This is a particularly relevant application, because the number of SOAP features can be increased systematically, and the very high number of resulting features is the main reason for the comparably high computational cost of the resulting regression models\cite{imba+18jcp,onat+20jcp,zuo+20jpcl}.
For all feature selection comparisons, models were trained on identical training sets of up to 11'854 environments, and errors reported for a separate test set of 1'317 environments.
We also provide a distinct example of the use of PCov-CUR feature selection for force field construction, training a feed-forward neural network based on Behler-Parinello atom-centred symmetry functions\cite{behl-parr07prl,behl11jcp}. We predict energy and forces for a data set containing 10,000 benzene configurations, including four different benzene polymorphs, and using 90\%{} of the structures for training and 5\%{} for validation and testing, respectively.
For all studies, details of the methods and data provenance can be found in the Supplementary Information.

\subsection{Training Set Selection}

We begin by investigating the use of PCov-CUR and PCov-FPS to select samples from a fixed training set. 
When choosing the most important training points out of a large pool of candidates, a fully-unsupervised scheme offers the clear advantage that one does not need to compute or measure the properties in advance, which is usually the time-consuming step.
However, one can often obtain inexpensively an approximate estimate of $\bY$, which could be used with PCov-based selection methods to reduce the number of accurate reference evaluations. 
Similarly, one may want to select samples from an existing training set to use them for more demanding data analytics, e.g. picking the most relevant snapshots from a molecular dynamics trajectory to use for feature selection or non-linear dimensionality reduction.

In \ref{fig:ts_lr_ly}, we show the error on a fixed randomly-selected test set of environments, for models trained on different sub-selections of the full train set. For each sub-selection, we train a linear ridge regression model
\begin{equation}
\bYhat = \bX (\bXs^T\bXs + \lambda \bI)^{-1} \bXs^T \bYs,
\end{equation}
with a regularisation $\lambda$ optimised by 2-fold cross-validation.
The curves in the figure compare the convergence with train set size for a random selection (the usual construction of a learning curve) with that obtained by different PCov-inspired methods. For each method, the shading indicates the range spanned as $\alpha$ is changed from 1 (indicated with full lines, equivalent to the standard unsupervised selection) to 0 (indicated with dashed lines, giving full weight to the supervised component). 

\begin{figure}
    \includegraphics[width=\linewidth]{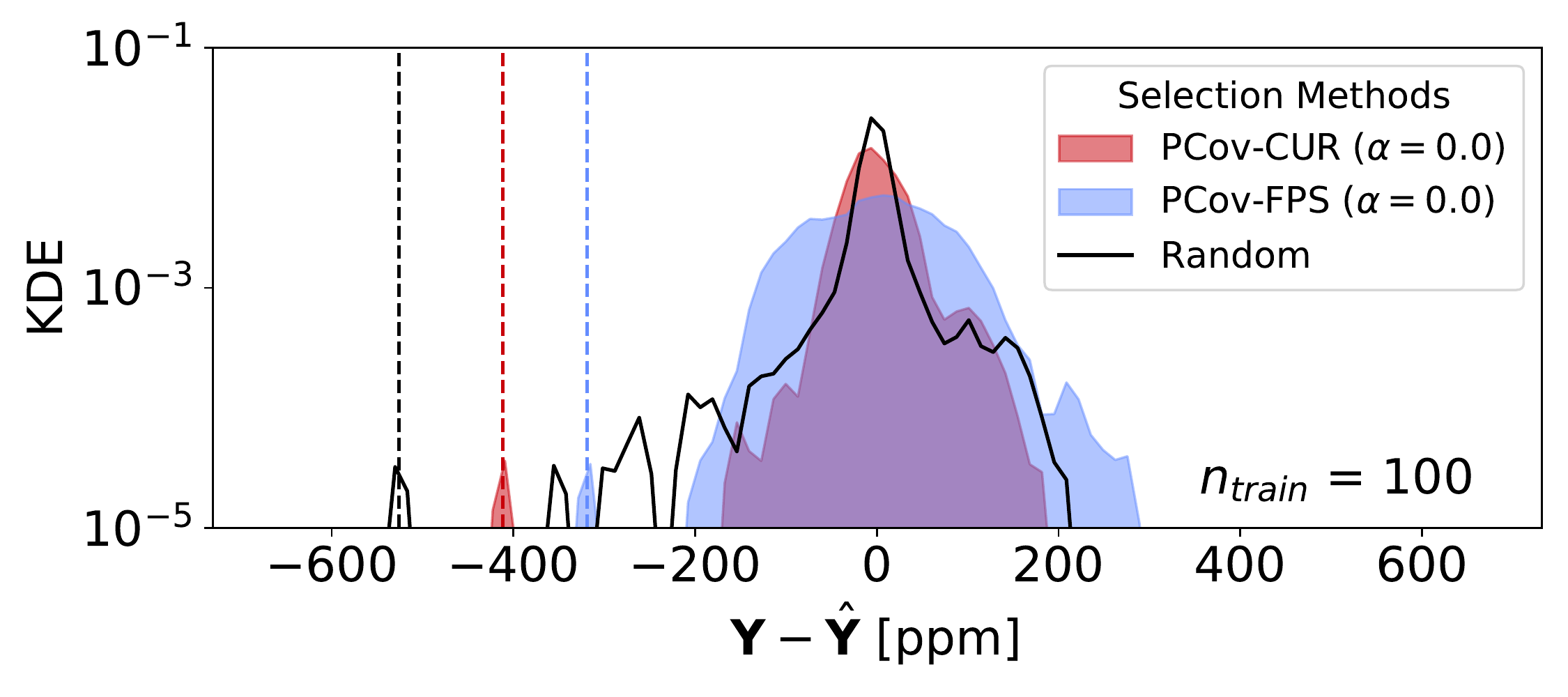}
    \caption{\textbf{Kernel density estimate (KDE) of errors} for random (black), PCov-CUR ($\alpha=0$, red) and PCov-FPS ($\alpha=0$, blue) selected training subsets, for $n_{train}=100$ (of 11'854). The dotted lines represent the maximum error for each respective training subset. Results for further $\alpha$ and $n_{train}$ are in Fig. S4.
    } 
     \label{fig:ts_csd_kde}
\end{figure}

\begin{figure}
    \includegraphics[width=\linewidth]{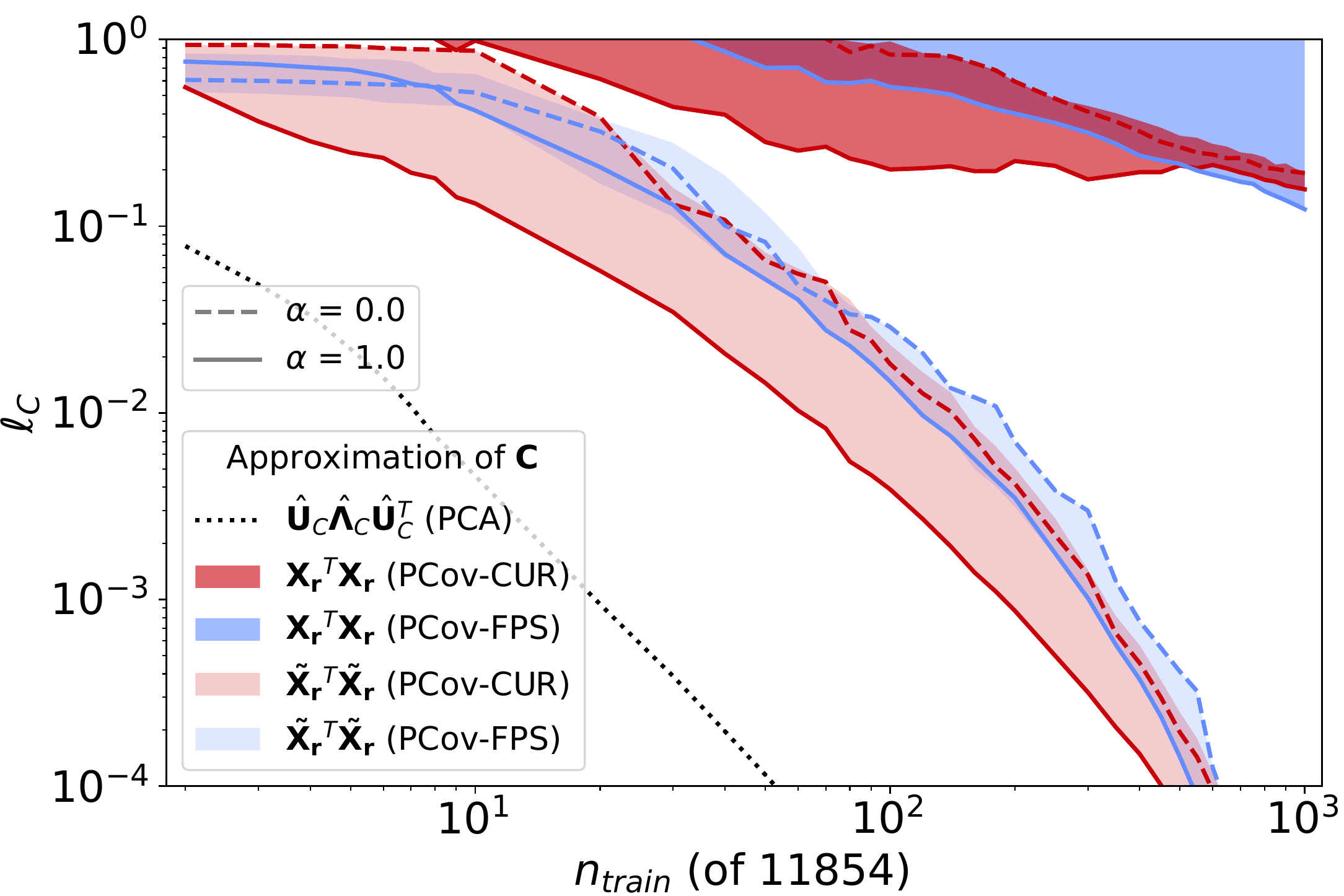}
    \caption{\textbf{Approximation of covariance matrix $\bC = \bX^T\bX$} using $\bXs$ and $\bXst$ for PCov-CUR and PCov-FPS on CSD-1000R. Shaded regions bound the minimum and maximum error across all $\alpha$ for each PCovR-inspired method. The dotted line corresponds to the approximation of $\bC$ from the $n_{train}$ eigenpairs corresponding to the largest eigenvalues.}
    \label{fig:ts_csd_covariances}
\end{figure}

Interestingly, when employing PCov-FPS, the unsupervised selection ($\alpha = 1$) performs best, usually being the only values that (marginally) improves upon a random selection, with errors increasing with decreasing $\alpha$. For PCov-CUR, the supervised and unsupervised methods perform comparably in terms of mean error, with a small decrease in error possible as $\alpha \to 0$.
To fully rationalise the performance of the different methods it is necessary to consider the error distribution, rather than just the mean error. 
As shown in \ref{fig:ts_csd_kde} the supervised limit of PCov-CUR reduces greatly the tails of the error distribution, indicating they provide more robust models, that are less susceptible to the presence of outliers in the test set. The PCov-FPS($\alpha=0$) selection results in a flatter distribution of errors, which has the smallest maximum error, but a larger mean error, as seen clearly in \ref{fig:ts_lr_ly}. 
In fact, one has to consider that in this example the test set is obtained by is randomly selecting environments from the same dataset that is used for training. Thus, a random sub-selection of the train set has exactly the same makeup as the test target. 
As a consequence, methods such as FPS and CUR are not guaranteed to match or improve upon the mean error compared to random selection. These observations are similar to those made in Ref.~\citenum{bart+17sa} for a kernel ridge regression model of the atomisation energy of small organic molecules.

\subsubsection*{Covariance-preserving sample selection}

Selecting a subset of the training structures, with and without incorporating a PCov component, also has an impact on unsupervised tasks. 
The reduced training set size potentially leads to a reduced rank of the covariance matrix computed from $\bXs$, $\bC_\br=\bXs^T\bXs$ but also to a skewed weighting of the importance of different features in the covariance built on the train set. 
This second effect can be mitigated by introducing a correction based on the matrix decomposition in Eq.~\eqref{eq:cur} to obtain a modified subset $\bXst$ which better preserves the sample covariance. In practice one computes
\begin{equation}
\tilde{\bC}_\br = \bXst^T\bXst = \bXs^T(\bXs^-)^T \bX^T \bX \bXs^-\bXs,
\end{equation}
which can then be diagonalised to compute a modified PCA projection matrix. 
If necessary, one could also evaluate explicitly the corrected feature matrix, 
\begin{equation}
\bXst =  \left[(\bXs^-)^T \bX^T \bX \bXs^-\right ] ^ {1/2} \bXs.
\label{eq:cur-xst}
\end{equation}
which can be useful if one wanted to perform feature selection based on a reduced train set. 

\begin{figure*}
    \centering
    \includegraphics[width=\linewidth]{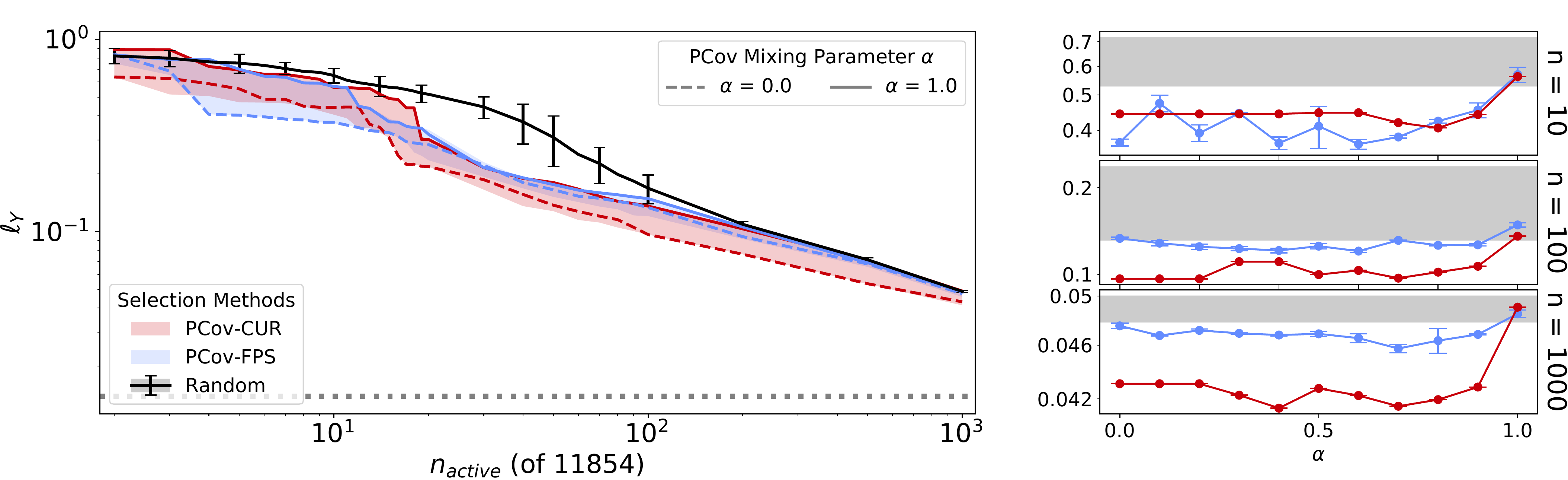}
    \caption{\textbf{Regression loss of CSD-1000R using sparse kernel ridge regression} with sample-selected active set. \legendexp{\lr} In the left panel, \gd{kernel}
    The right side panels show the performance at $\ns=10, 100, 1000$ (of 11'854).
    }
    \label{fig:ss_skrr}
\end{figure*}

\ref{fig:ts_csd_covariances} shows the error in reproducing the covariance matrix $\bC = \bX^T \bX$ with reduced sample sets. We can compare these results to the approximation of $\bC$ by its eigendecomposition $\hat{\bC} = \bUhat_C \bLAMhat_C \bUhat^T$, computed with the top $\ns$ eigenvalues and their corresponding eigenvectors. Computing the covariance using the sample set $\bXs$ results in large error ($\ell_{C}\approx 0.3$ with 1'000 samples), with convergence only as $\ns$ approaches $\nf$ (2520 in this case). 
Computing the covariance with $\bXst$ reduces the loss considerably, with a covariance loss of $\ell_{C} = 0.01$ possible with as few as 60 / 11'854 of the training points. PCov-CUR typically outperforms PCov-FPS, and for both the approximation degrades as $\alpha \to 0$.

\subsection{Active Set Selection for Sparse Kernels} \label{sub:active-set}

Another context in which one wants to perform sample selection is when building a sparse kernel model, using the projected-process approximation\cite{carledwardrasmussen2005}. In sparse kernel ridge regression one defines an ansatz for the properties of a sample as 
\begin{equation}
\by(\bx) \approx \sum_{i\in M} \mathbf{w}_i k(\bx, \bx_i),
\end{equation}
where $M$ indicates a set of ``reference samples'' that effectively constitute a basis to expand $\by(\bx)$.
The weights ${\mathbf{w}_i}$ are optimised to minimise the regression error on the train set.
In the projected-process approximation, the reference samples $\bXs$ (also known as ``active points'') corresponds to a sample-selected subset of the training set. 
If we denote $\bKNN$ as the kernel matrix between a dataset, described by the feature matrix $\bX$, and itself; $\bKNM$ as the kernel between $\bX$ and $\bXs$;  $\bKMM$ as the kernel computed between $\bXs$ and itself, the values of the predicted properties for the train set are given by 
\begin{equation}
    \bYhat = \bKNM  (\bKNM^T \boldsymbol{\Lambda}^{-1} \bKNM + \bKMM)^{-1} \bKNM^T \boldsymbol{\Lambda}^{-1} \mathbf{Y},
    \label{eq:skrr}
\end{equation}
where $\boldsymbol{\Lambda}$ is a regularisation matrix, that is usually taken to be simply a scalar, in which case \eqref{eq:skrr} reduces to that reported in Ref.~\citenum{helf+20mlst}. 
In this example, we use the radial basis function (RBF) kernel $k(\bx, \bx^\prime) = \exp\left(-\gamma \lVert \bx - \bx^\prime \rVert^2\right)$, and we choose $\gamma = 10^{-4}$, which optimises the combined regression and reconstruction loss for the full feature vectors (see Fig. S5A).

\ref{fig:ss_skrr} compares the performance of a sparse kernel ridge regression as a function of the number of reference points, selected by different techniques. All data-driven selection methods outperform by up to a factor of two the random baseline, particularly in the small active-set limit. 
CUR marginally outperforms FPS at all but the smallest active set sizes, and PCov-inspired methods generally perform better than their unsupervised counterparts, with an error decreasing further as $\alpha \to 0$. 

The sample selection that underlies ~\ref{fig:ss_skrr} is the same as for the train set selection in the previous Section, i.e. based on a linear PCovR framework based on the raw $\bX$ features. The fact that the representative samples chosen with a linear framework are effective for a non-linear kernel model underscores the robustness of the selection criteria, and is important as in most cases one wants to use a fixed sub-selection while tuning the model, e.g. by optimising hyperparameters or testing different kernels. 
If one wanted to select an active set that is consistent with the kernel-induced metric, it would suffice to substitute the kernel matrix to the matrix of scalar products in the definition of the sample distance Eq.~\eqref{eq:fps-K} or the leverage scores Eq.~\eqref{eq:leverage-K}.

\begin{figure}[!ht]   
    \begin{subfigure}[b]{\linewidth}
    \caption{\textbf{Global feature reconstruction error (GFRE)}}
    \label{fig:fs_csd_gfre}
    \includegraphics[width=\linewidth]{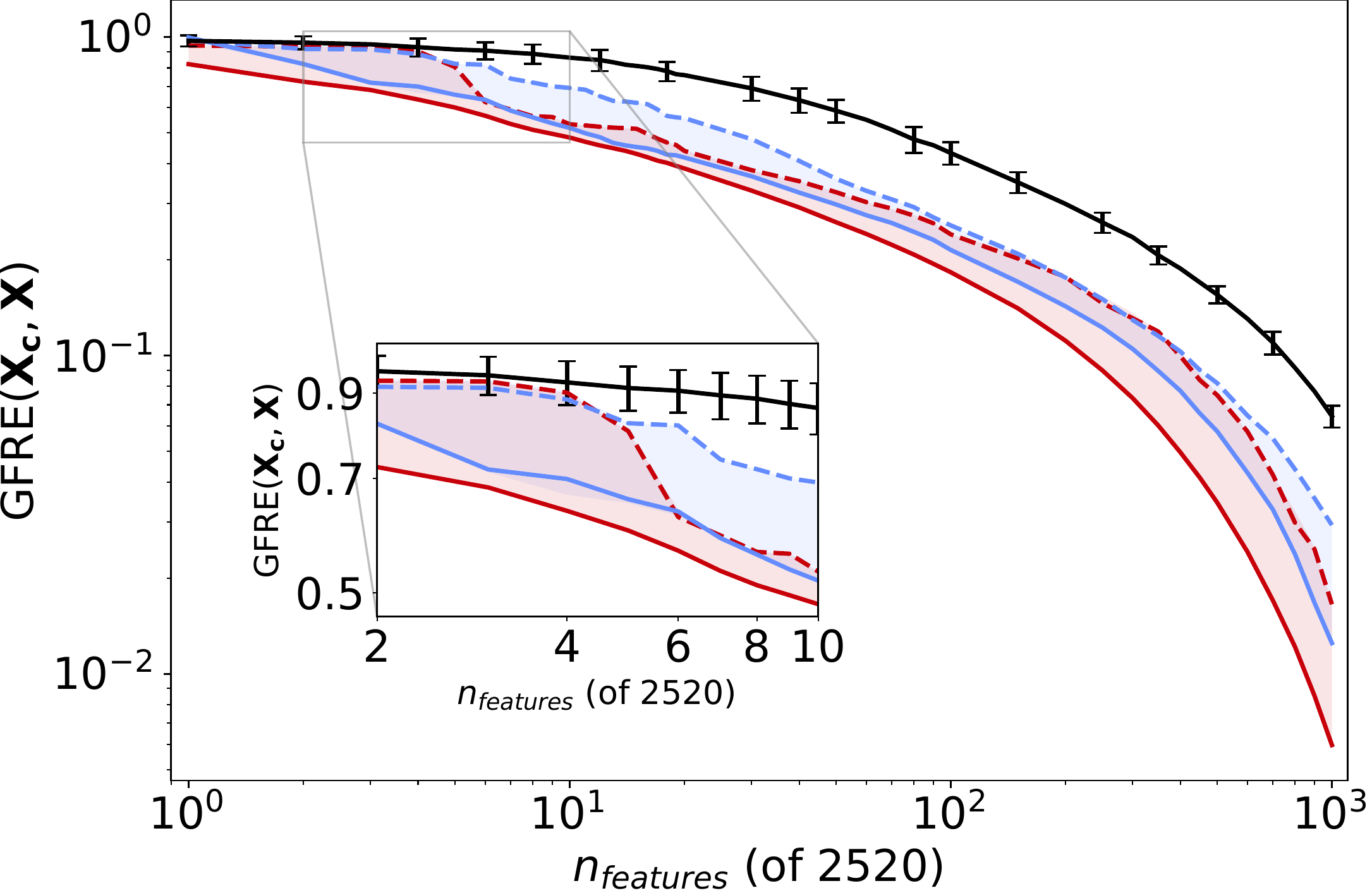}
    \end{subfigure}
    \begin{subfigure}[b]{\linewidth}
    \caption{\textbf{Approximation of the gram matrix $\mathbf{K} = \mathbf{XX}^T$} }
    \includegraphics[width=\linewidth]{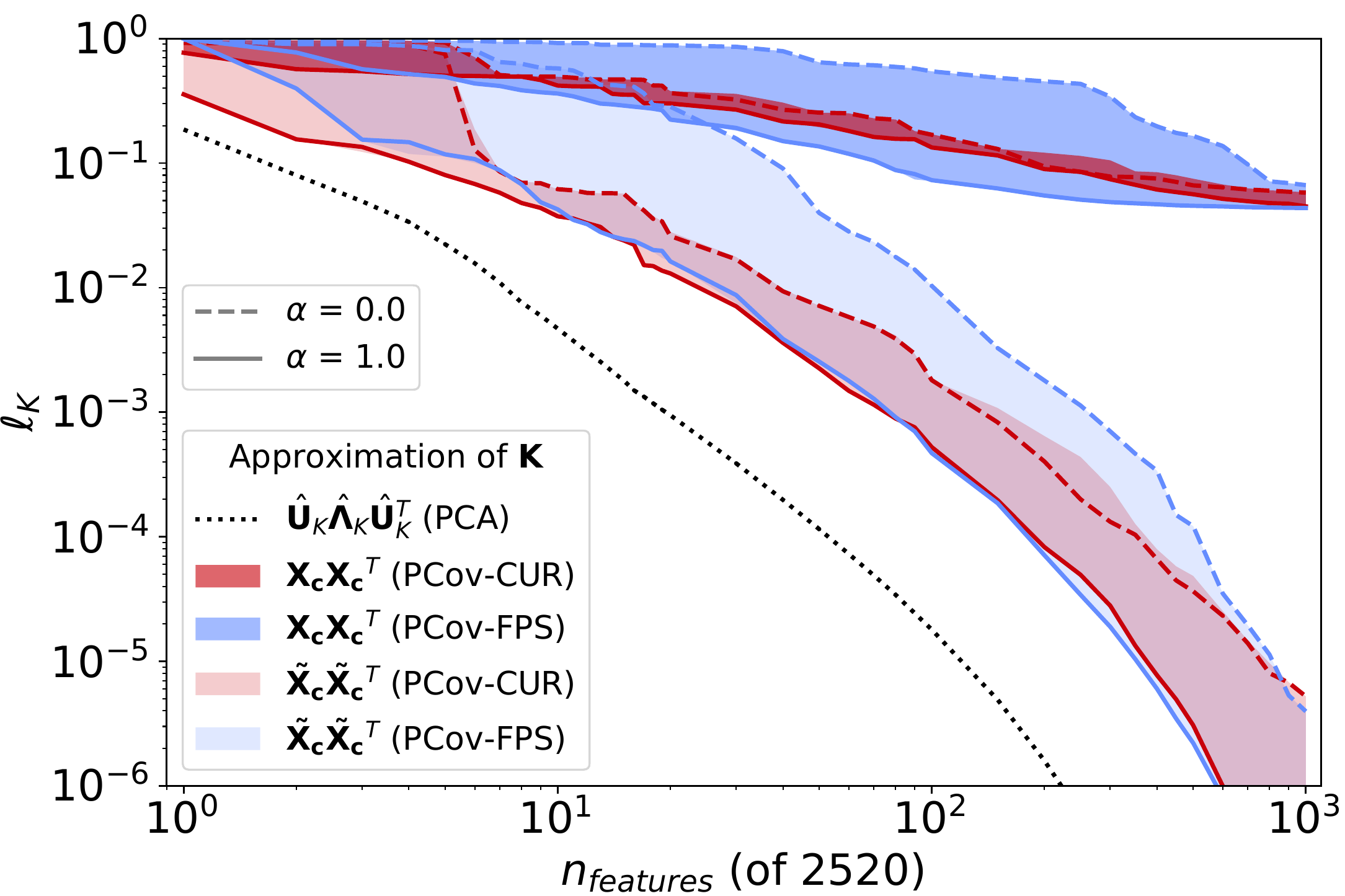}
    \label{fig:fs_csd_pca_main}
    \end{subfigure}
    \caption{\textbf{Performance of feature subsets of CSD-1000R in unsupervised tasks.} \legendexp{\lc} }
    \label{fig:fs_csd_unsup}
\end{figure}

\subsection{Feature Selection}
\label{sec:fs}

The very same techniques that can be applied to select the most important samples can be used to identify the features that provide most information about the training data set, and -- with a PCov component -- about structure-property relations. 
Selecting the most relevant features is useful because the cost of evaluating $\bx$ usually scales with $\nf$, and the cost of evaluating a model built on $\bx$ similarly increases with the size of the feature vector. 

\subsubsection*{Global Feature Reconstruction Error (GFRE)}
An obvious criterion to assess the performance of a feature-selection scheme is to verify whether the chosen subset of features contains comparable amount of information to the full feature vector. 
A quantitative measure of the relative information content of two sets of features is given by the recently-introduced \textit{global feature-space reconstruction error} (GFRE)\cite{gosc-2020}, that estimates it in terms of the error one incurs when learning one set from another: 
\begin{equation}
\operatorname{GFRE}(\bX, \bX^\prime) = \sqrt{\left\|\bX'-\bX\bP{X}{X'}\right\|^2/\ns}.
\end{equation}
The reconstruction error is evaluated on the test set, the linear projection $\bP{X}{X'}=(\bX ^T \bX + \lambda \bI)^{-1} \bX^T \bX^\prime$ is computed on the training set, and $\lambda$ is an appropriate regularisation parameter, determined by cross-validation. Both $\bX$ and $\bX'$ are taken to be standardised. 
Fig.~\ref{fig:fs_csd_gfre} shows that the GFRE decreases rather slowly with the number of selected features: this is to be expected as SOAP power spectrum features are linearly independent, which in combination with a diverse dataset containing many different types of chemical environments leads to a high intrinsic dimensionality of the feature space. 
In  both the small and large $\nf$ limit, using a PCov-augmented scheme with $\alpha<1$ leads to a degradation of the GFRE, while for intermediate $\nf$, the effect of $\alpha$ is small. 
This is unsurprising, because the GFRE reflects only the information content of the feature vectors, and not the regression accuracy: the $\alpha=1$ case is the most compatible with the goal of minimising the error in reconstructing the full feature vectors. 
Despite this fact, all data-driven selection schemes systematically reduce the GFRE compared to a random selection, including in the $\alpha\rightarrow 0$ limit. CUR generally outperforms FPS -- although it is more sensitive to an increase in the weight given to the supervised component.

\subsubsection*{Distance-preserving feature selection}

The calculation of the GFRE incorporates a linear transformation of the selected features. When one wants to use $\bX_\bc$ in the context of unsupervised-learning algorithms, that depend on preserving the value of the scalar products -- and hence the distances -- between feature vectors, it necessary to incorporate a similar linear transformation, analogous to that discussed for the case of sample selection. 
To see how such transformation would help, imagine the case in which two features are identical. Dropping one would entail no information loss, but would distort distances in feature space, de-emphasising the component that has been discarded. Scaling the retained feature by $\sqrt{2}$ would restore exactly the original metric.

Irrespective of how the feature selection is performed, one can use the matrix decomposition in Eq.~\eqref{eq:cur} to obtain a modified subset $\bXft$ which better preserves the Euclidean distances in feature space. %
We start with the approximation of $\bX$ using the typical CUR decomposition, where
\begin{equation}
    \bXhat \approx \bXf \left(\bXf^- \bX \bXs^-\right) \bXs.
\end{equation}
\begin{figure*}
   \includegraphics[width=\linewidth]{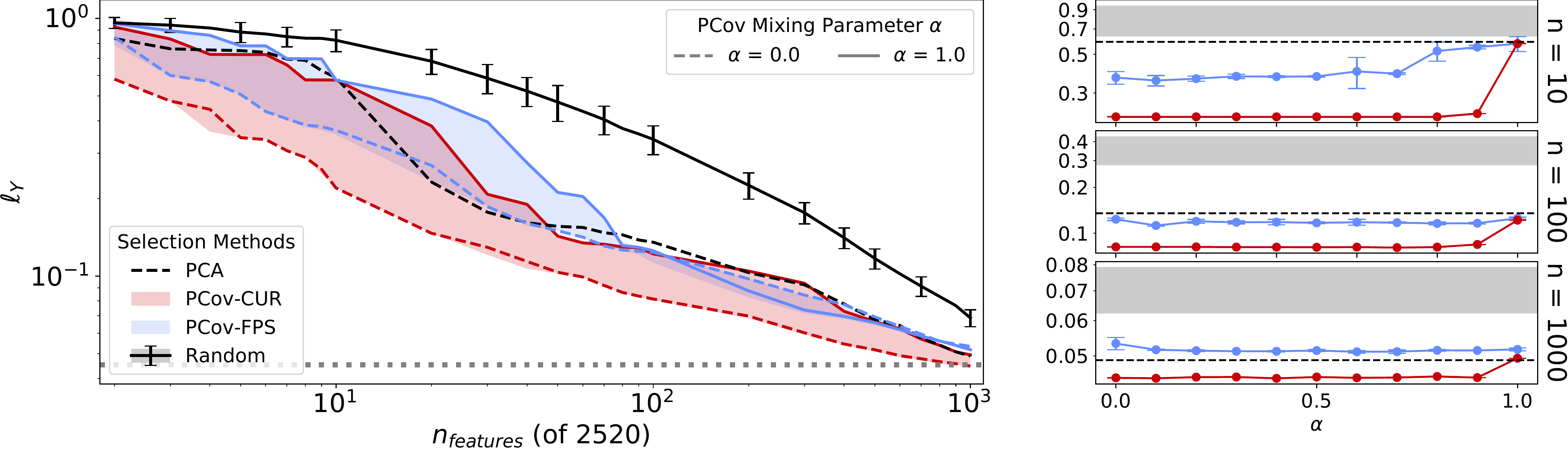}
    \caption{\textbf{Ridge regression losses for different feature selection metrics for CSD-1000r} \legendexp{\lc} In the left panel, \gd{feature vectors} The black dotted line denotes the regression loss for an analogous PCR, where the $\ns \times \nf$ latent-space projection obtained from PCA is used to train the regression model. Side panels depict losses for various $\alpha$ values used to select 10, 100, and 1000 (of 2520) features.
    }
    \label{fig:fs_csd_lr}
\end{figure*}
We can determine $\bXft$ by assuming $\bXs = \bX$ and constructing $\bXft$ such that $\bXft \bXft^T \approx \bX\bX^T$,
\begin{equation}
\bXft = \bXf \left[\bXf^- \bX \bX^T (\bXf^-)^T\right ] ^ {1/2}.
\label{eq:cur-xft}
\end{equation}
The $\nf \times\nf$ matrix $\left[\bXf^- \bX \bX^T (\bXf^-)^T\right ] ^ {1/2}$ can be computed once and re-used every time one needs to compute $\bXft$, even with out-of-sample data. 
Fig. \ref{fig:fs_csd_pca_main} shows the error in reproducing the Gram matrix $\bX\bX^T$ with the selected features, as a measure of the distortion in feature-space metrics. 
The error one incurs when truncating a PCA latent space is by construction the minimal that one can achieve with a linear $\nf$-dimensional projection of $\bX$. 
Feature selection leads to a large distortion of the underlying metric, with $\ell_{K}\approx 0.1$ even with more almost 50\%{} of the features included in $\bXf$. 
The use of a correction to the selected feature matrix, as in Eq.~\eqref{eq:cur-xft}, improves dramatically the accuracy in preserving the feature-space metric, even though asymptotically a PCA projection outperforms selected columns by up to a factor of 10. The use of a hybrid, PCov-like selection does affect the accuracy of the reconstruction of the Gram matrix, with PCov-FPS generally being more severely affected by taking $\alpha\rightarrow 0$ than PCov-CUR.

\subsubsection*{Linear Ridge Regression}

The advantages of using PCov-augmented feature selection are clearest when considering their application to regression tasks.
To assess the performance of a given selection scheme, we consider the error in approximating a target $\bY$ given a feature subset $\bXf$
\begin{equation}
\bYhat = \bXf(\bXf^T\bXf + \lambda \bI)^{-1} \bXf^T \bY
\end{equation}
The results in \ref{fig:fs_csd_lr} demonstrate that both PCov-CUR and PCov-FPS improve the regression performance when compared to random selection, with comparable losses often achieved with 10 times fewer features. PCov-CUR generally out-performs PCov-FPS. As shown in the side panels, in the case of CUR the performance improves as $\alpha \to 0$, with performances being near constant for $\alpha < 1$, and the reference, full-features accuracy being reached with $\nf\approx 1000$. The same is true for FPS-based selection up to $\nf\approx 100$, after which the value of $\alpha$ has little effect, and the lowest error is observed for intermediate values of $\alpha$. 

We also include the results obtained from principal components regression, where the latent space projection from a PCA with $\npca=\nf$ is used in place of $\bXf$ to predict the materials properties. PCA provides the best unsupervised approximation of the feature vector, and so it serves as a baseline to assess the improvements that can be obtained incorporating a supervised component to feature selection. 
Indeed, regression based on the principal components used as features usually performs better than unsupervised FPS, and comparably to unsupervised CUR. PCov-augmented selections, instead, consistently out-perform those from PCA -- which evidences that retaining the largest variance components, as one does in PCA, does not necessarily yield features that are predictive for the properties of interest\cite{jolliffe_note_1982}, which is also relevant for the methods that rely on feature (co)variance to construct a hierarchy of increasingly complex representations of the atomic structure\cite{niga+20jcp}.

\begin{figure*}
    \includegraphics[width=\linewidth]{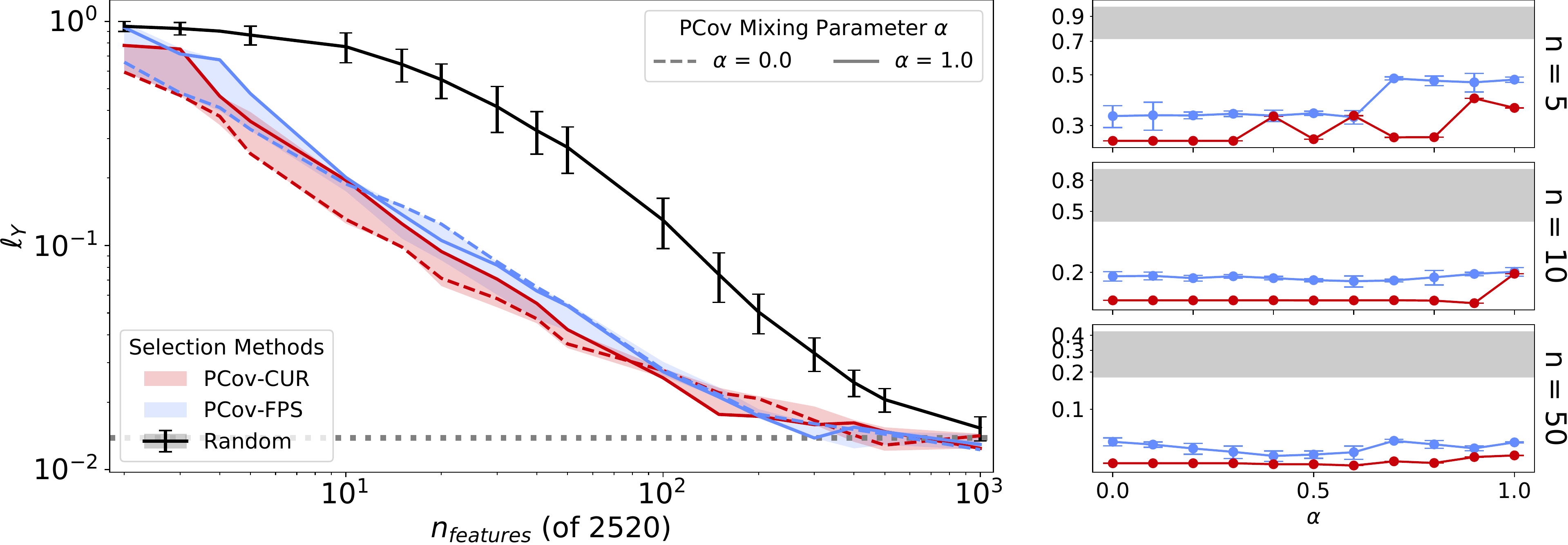}
    \caption{\textbf{Results using low-rank kernels built on selected features.} \legendexp{\lc} In left panel, \gd{feature vectors} The side panels shows the KRR loss for separate selections of 5, 10, and 50 (of 2'520) features for different values of $\alpha$.}
    \label{fig:fs_csd_krr}
\end{figure*}

\subsubsection*{Kernel Ridge Regression}
\label{sec:kernels}

The feature-selected $\bXcol{\bc}$ can also be used to compute an approximation $\hat{\bG}$ of the kernel matrix, by simply using the compressed feature vectors to evaluate the (non-linear) kernel function $\hat{K}_{ij}=k((\bX_\bc)_i,(\bX_\bc)_j)$.
As in Section~\ref{sub:active-set}, we use an RBF kernel with $\gamma=10^{-4}$.

We assess the performance of the approximate kernel by fitting a kernel ridge regression model
\begin{equation}
    \bYhat \approx \mathbf{\hat{K}} (\mathbf{\hat{K}} + \lambda \bI)^{-1} \bY,
    \label{eq:loss-fs-lr}
\end{equation}
and computing the error on the test set. 
The non-linearity in the definition of $k$ means that regression is performed in a different feature space than $\bX$, which improves by a factor of four the regression loss with respect to a linear model based on the full $\bX$; however the rationale underlying unsupervised and PCov-augmented selection methods is partly inconsistent with the ultimate measure of success.

Nevertheless, as shown in \ref{fig:fs_csd_krr}, kernels built on a subset of the features chosen by a PCov-FPS or PCov-CUR method outperform those based on a random selection of equivalent size by up to an order of magnitude, and match a kernel built on the full $\bX$ with just $\nf\approx 400$. 
With increasing number of features, the value of $\alpha$ has a smaller effect than in the case of linear regression.

\subsubsection*{Feature Selection for Neural Network models}
\label{sub:nn}

Thus far we have demonstrated the effects of semi-supervised feature and sample selection for simple ML models, with a deterministic relationship between features, training set, and test error. %
More complex models, such as those based on artificial neural networks (NN) can reproduce an arbitrary, non-linear dependence of the target properties $\bY$ on the input features $\bX$.
NNs are ``trained'' by iterative minimisation of the ($L^2$) loss between the NN output ${\bY}_{\textrm{NN}}$ and reference values ${\bY}$ with respect to the free parameters in the network, usually called NN weights.
After training the NN can be used to predict $\bY$ for an arbitrary $\bX$. 
An example of a NN framework that has become commonplace in atomistic modeling is that introduced by Behler and Parinello to describe interatomic potentials\cite{behl-parr07prl,behl11jcp, behl11pccp}, which is based on the decomposition of atomic configurations and total energies into local, atom-centred environments and associated energy contributions.
Environments are described using two-body and three-body symmetry functions (SF), which correspond to a projection of two and three body correlations between the neighbours of the target atomic centre on a bespoke non-orthogonal basis. %
These constitute the features $\bX$, which are used as the input layer, which is connected to one or more narrower, fully connected ``hidden layers'', and finally combined to predict an atom-centred decomposition of the potential energy of the system.
Nodes are linked via non-linear activation functions $f_a$ and each node $i$ in layer $k-1$ is connected to each node $j$ in layer $k$ with a tunable weight $w_{ij}^k$.
For a single hidden layer with $N$ nodes a Behler-Parinello NN can be expressed as
\begin{equation}
    Y_{\textrm{NN}} = f_a^2 \left[ u^2 + \sum_{n=1}^{N} w_{n1}^1 f_a^1 \left( v_{n}^1 + \sum_{i=1}^{f} w_{i n}^1 X_{i} \right) \right]
\end{equation}
where $u^k$ and $v_{j}^k$ are adjustable offsets.
Besides the more complicated functional relationship between features and properties, BPNN differ from the other examples we consider here because they include the derivative of the energy with respect to atomic coordinates (the forces) as part of the regression targets.

\begin{figure*}
    \begin{subfigure}[b]{0.45\linewidth}
    \includegraphics[width=\linewidth]{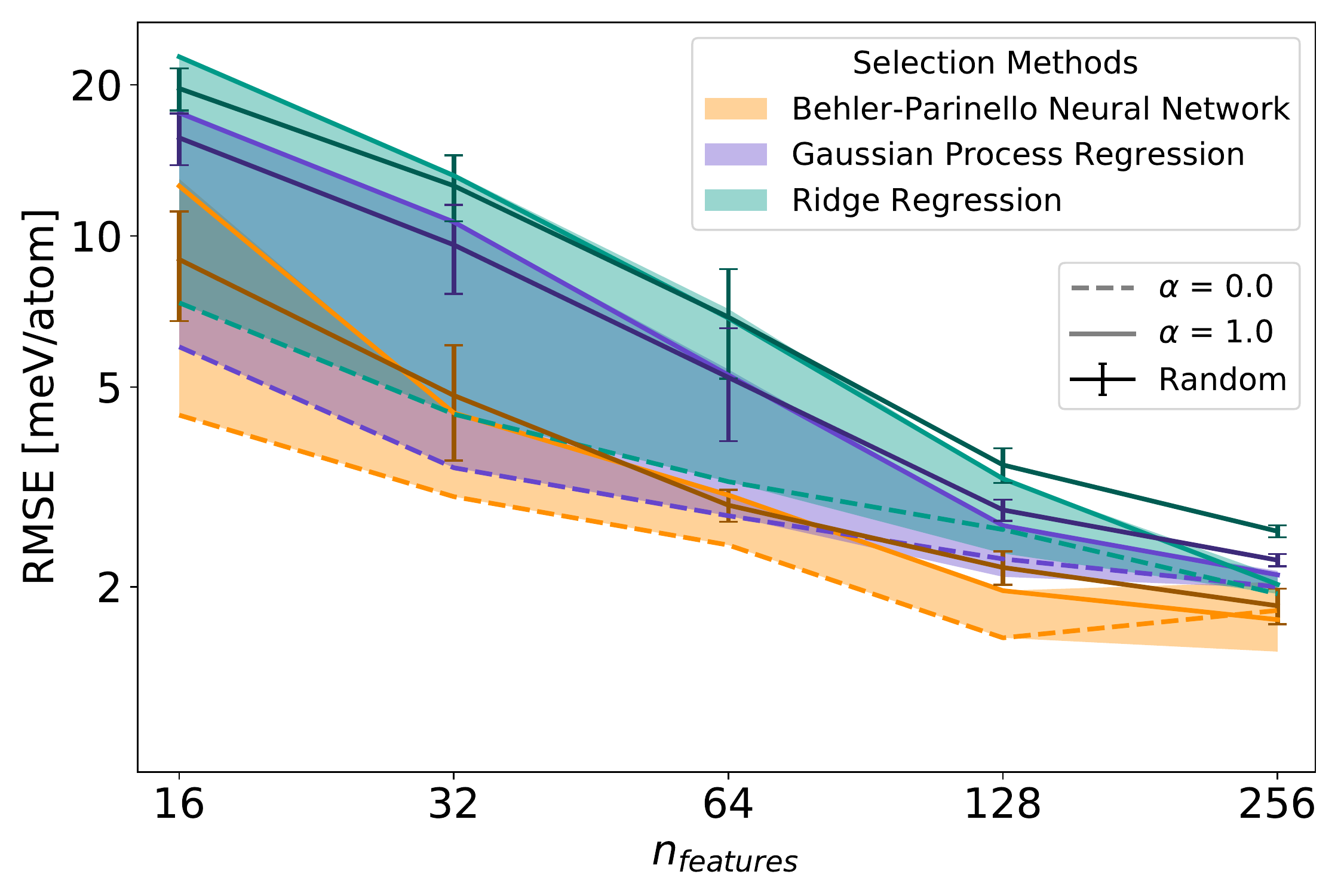}
    \caption{\textbf{RMSE in energies} using Ridge Regression (green), Gaussian Process Regression (purple), and BP NN (gold). }
    \label{fig:all_energy}
    \end{subfigure}\begin{subfigure}[b]{0.45\linewidth}
    \centering
    \includegraphics[width=\linewidth]{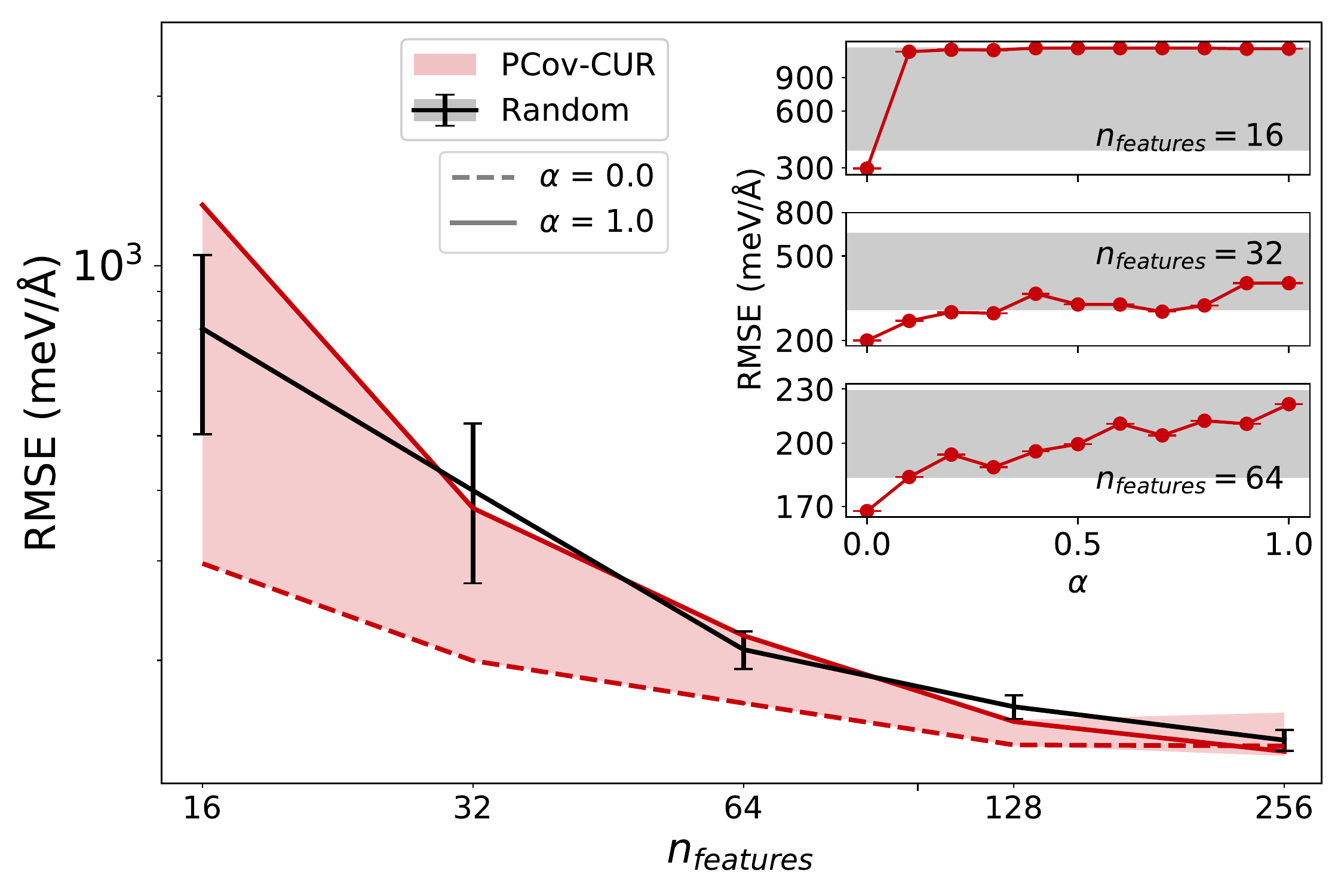}
    \caption{\textbf{RMSE in forces} for BP NN. Insets denote the dependence on $\alpha$ for 16, 32, or 64 features.}
    \label{fig:NN_forces}
    \end{subfigure}
    \caption{\textbf{Root-Mean-Squared-Errors (RMSE) for energies (left) and forces (right) from various models and feature selections}. The dotted and solid lines denote results for $\alpha = 0$ and $\alpha = 1$, respectively. Darker error bars and / or grey regions represent the mean and standard deviation of errors for random feature selection.}
    \label{fig:NN-energies}
\end{figure*}

The evaluation of SFs is computationally demanding, and the number that would be generated by systematically spanning the parameters that define their functional form increases quadratically or cubically for two and three-body SFs. 
Furthermore, Behler-Parrinello symmetry functions are not orthogonal and are often highly redundant, which together with considerations of computational efficiency motivates strongly the selection of a small number of SFs. 
Traditionally, parameters have been optimised with a combination of chemical intuition and trial-and-error, but more recently it was shown that CUR and FPS provide a viable strategy to choose a small set of SFs out of a large pool of candidates\cite{imba+18jcp}. 
To investigate how a PCov augmentation impacts the selection of features in the context of a NN potential, we consider a database of 10'000 molecular structures corresponding to thermally-distorted benzene crystals, and construct a feature matrix that consist in 452 two and three-body BP symmetry functions, by varying the functions parameters on a regular grid, following the protocol in Ref.~\citenum{imba+18jcp}. We focus on PCov-CUR, that has been shown to yield consistently better performance than PCov-FPS, albeit with larger computational effort. 

We begin by discussing how the non-linear nature of the NN model affects the performance of the feature selection schemes. We compare linear regression, kernel ridge regression using an RBF kernel, and the Behler-Parrinello NN, using as inputs subsets of the entries of a full feature vector containing 452 symmetry functions, determined by PCov-CUR and random selection. Each model was trained using only energy information, cross-validated, and with errors reported for identical training, validation, and testing samples sets, respectively.
For each selection of input features we train 4 NN potentials, using a random 90:5:5 train:validation:test splitting of the dataset and weight initialisation, and solely energy in the fit. We regularise the regression using an early-stopping criterion, and report for each training exercise the best out of the four results.

\ref{fig:NN-energies} demonstrates the interplay between the nature of the features, the selection protocol, and the regression model. 
Models that incorporate increasing levels of non-linearity lead to better regression performance, with KRR usual outperforming LR by 25\%{}, and NN reducing the energy RMSE by an additional 40\%{}. 
A side-effect of using features that are highly correlated with each other, is that the performance of a random selection is not particularly poor, in comparison with unsupervised CUR subset. 
A PCov-CUR($\alpha=0$) selection that incorporates target information, however, allows one to identify the features that are most relevant for the construction of the potential, which determines a very substantial reduction of the energy RMSE, particularly for small $\nf$ values. 
Irrespective of the number of selected SF, PCov-CUR selections of SFs consistently outperform (and never perform worse than) the average random selection.
Further, PCov-CUR selections of SF, which are biased towards correlating with the target property ($\alpha = 0$), consistently outperform unsupervised CUR selections of SF, with similar improvements being observed for all regression schemes.
As the number of selected SF approaches the full set, the difference between a supervised and unsupervised feature selection decreases, but can result in an improvement over a random selection even for $\nf=256$.

This improvement is particularly remarkable because of the gap between the linear, energy-based supervised framework that underlies the PCov augmentation, and the non-linear predictions of atomic forces that is assessed in the figure, which indicates that the methods we propose are robust and can be applied to improve feature selection beyond linear or kernel methods. 
Overall, the best PCov-CUR selection allows to reduce by 50\%{} the number of SF while retaining roughly the same force RMSE. This results into direct computational savings when using the NN potential, and into a simpler, less memory-intensive task when training the model.

\section{Conclusions}

Selecting the samples and/or features that are most relevant for a desired task out of a large pool of candidates can be very advantageous from a computational point of view, and helps revealing the most important, or insightful descriptors. 
This is particularly useful in cases in which features can be constructed in a systematic way, leading potentially to large and redundant input representations. 
Unsupervised methods, based on a low-rank approximation of the feature matrix, or on maximising the diversity of samples or features, provide an effective approach to prune a training set or a collection of descriptors with little loss in model performance. 

Whenever the end goal of the featurization is to serve as the input of a supervised model, it is appealing to incorporate the regression target into the feature selection, which we do here by combining two methods (FPS and CUR) with a semi-supervised linear scheme, PCovR. 
For a variety of different problems, ranging from reducing the size of a training set, to active point selection in sparse kernel regression, to linear and non-linear model fitting, we find that such PCov-augmented selections outperform almost universally their unsupervised counterparts, which makes it possible to obtain, in practice, models that achieve comparable prediction accuracy to the model based on full features and training set, but at much reduced effort.
The simplicity of PCov-FPS and PCov-CUR, the ease by which they can be extended to incorporate non-linearity, and the empirical evidence showing that they can also improve the accuracy of non-linear, neural network models, make them applicable to virtually any regression use case. This, together with the availability of an open source implementation\cite{skcosmo}, gives these methods the potential to become a standard tool in the application of data-driven methods to different fields of science.

\section{Acknowledgements}
MC, RKC, BAH acknowledge the ERC Horizon 2020 grant no. 677013-HBMAP. EE acknowledges support from Trinity College, Cambridge and compute time from the Swiss National Computing Centre (projects s960, s1000).

\section{Code Availability}
Software for computing feature or sample subsets using PCov-CUR or PCov-FPS can be found at \url{www.github.com/cosmo-epfl/scikit-cosmo} with documentation at \url{scikit-cosmo.readthedocs.io}.

\appendix

\section{PCov feature space}
\newcommand{\bLt}{{\tilde{\mathbf{L}}}}
Given the singular value decomposition of the feature matrix $\bX = \bU_\bG \bLAM^{1/2} \bU_\bC^T$ it is possible to define a PCov feature matrix $\bXt = \bU_\bG \bLt^{1/2} \bU_\bC^T$, with
\begin{equation}\label{eq:pcov-features}
\bLt = \alpha \bLAM + (1-\alpha) \bU_\bC^T\bY\bY^T\bU_\bC.
\end{equation}
It is easy -- albeit tedious (see Sec. S1) -- to check that $\bXt\bXt^T = \bGt$ and $\bXt^T\bXt = \bCt$. In general, for $\alpha\ne 1$, the matrix $\bLt$ is not diagonal, and so the singular vectors of $\bX$ are not given by $\bU_\bG$ and $\bU_\bC^T$, but can be obtained by diagonalising $\bGt$ or $\bCt$. 
Eq.~\eqref{eq:pcov-features} also makes it possible to diagonalise $\bLt = \bU_\bLt \tilde{\bLAM} \bU_\bLt^T$ and compute $\bU_{\bGt} = \bU_\bG\bU_\bLt$ and $\bU_{\bCt}=\bU_\bC\bU_\bLt$,

\end{document}